\documentstyle[12pt,cite,epsf]{article}
\renewcommand{\thefootnote}{\fnsymbol{footnote}}
\def \be{\begin{equation}}
\def \ee{\end{equation}}
\def \bea{\begin{eqnarray}}
\def \eea{\end{eqnarray}}
\def \bem#1{\renewcommand{\thefootnote}{\arabic{footnote}}\footnote{#1}}
\def \bm{\boldmath}

\def \branch{\mathrm {Br}\,}

\def \ea{{\it et al.\/}}

\def \eq#1{Eq.~(\ref{#1})}
\def \eqs#1#2{Eqs.~(\ref{#1})--(\ref{#2})}

\def \figs#1#2{Figs.~\ref{#1}--\ref{#2}}

\def \GeV{{\mathrm{GeV}}}
\def \ha{Hamiltonian}

\def \Im{{\mathrm{Im}}\,}

\def \MeV{{\mathrm{MeV}}}

\def \nnu{\nonumber}

\def \Re{{\mathrm{Re}}\,}
\def \rf{Ref.~\cite}

\def \T{{\it T\/}}

\def \Vec#1{\mbox{\bf{#1}}}
\def \fba{A_{\mathrm {FB}}(\sh)}

\def \rhadroncc{R_{\mathrm {had}}^{c\bar{c}}(\sh)}
\def \rhadron{R_{\mathrm {had}}}
\def \rcont{R_{\mathrm {cont}}}
\def \rcontcc{R_{\mathrm {cont}}^{c\bar{c}}(\sh)}
\def \rrescc{R_{\mathrm {res}}^{c\bar{c}}(\sh)}
\def \lop{P_{\mathrm L}}
\def \trp{P_{\mathrm T}}
\def \nop{P_{\mathrm N}}
\def \eN{\Vec{e}_{\mathrm{N}}}
\def \eL{\Vec{e}_{\mathrm{L}}}
\def \eT{\Vec{e}_{\mathrm{T}}}
\def \cseff {c_7^{\mathrm{eff}}}
\def \ceff {c_9^{\mathrm{eff}}}
\def \c9eff*{c_9^{\mathrm{eff}*}}
\def \a{\alpha}

\def \g{\gamma}
\def \G{\Gamma}

\def \l{\lambda}

\def \m{\mu}
\def \n{\nu}
\def \p{\pi}

\def \t{\tau}
\def \mh{\hat{m}}
\def \mlh{\hat{m}_l}
\def \mbh{\hat{m}_b}
\def \mch{\hat{m}_c}
\def \msh{\hat{m}_s}
\def \mth{\hat{m}_{\t}}
\def \sh{\hat{s}}

\def \today{\ifcase\month\or
  January\or February\or March\or April\or May\or June\or
  July\or August\or September\or October\or November\or December\fi
  \space\number\year}
\newcounter{Section}
\def \theSection{\Roman{Section}} 
\newcommand{\Sec}[1]{\refstepcounter{Section}%
\centerline{\bf\theSection. #1}\setcounter{equation}{0}}
\renewcommand{\theequation}{\arabic{Section}.\arabic{equation}}
\newcounter{Subsection}[Section]

\newcounter{Subsubsection}[Section]

\newcounter{Appendix}

\newcommand{\app}[1]{{\addtocounter{Appendix}{1}}%
\centerline{\bf APPENDIX: #1}\setcounter{equation}{0}%
\renewcommand{\theequation}{\Alph{Appendix}\arabic{equation}}}

\def\ib#1#2#3{{\it ibid.\/}~{\bf#1} (19#2) #3}

\def\np#1#2#3{{\it Nucl.~Phys.\/}~{\bf B#1} (19#2) #3}
\def\pl#1#2#3{{\it Phys.~Lett.\/}~{\bf B#1} (19#2) #3}
\def\pp{{\it preprint\/} }
\def\prd#1#2#3{{\it Phys.~Rev.\/}~{\bf D#1} (19#2) #3}
\def\prl#1#2#3{{\it Phys.~Rev.~Lett.\/}~{\bf #1} (19#2) #3}
\def\prp#1#2#3{{\it Phys.~Rep.\/}~{\bf #1} (19#2) #3}

\def\sov#1#2#3{{\it Sov.~J.~Nucl.~Phys.\/}~{\bf #1} (19#2) #3}
\def\yadfiz#1#2#3{{\it Yad.~Fiz.\/}~{\bf #1} (19#2) #3}
\def\zpc#1#2#3{{\it Z.~Phys.\/}~{\bf C#1} (19#2) #3}
\begin{document}
\begin{flushright}
PITHA 96/4 \\ hep-ph/9603237 \\ March 1996
\end{flushright}
\begin{center}
\LARGE \bf Lepton Polarization in the Decays \bm{$B\to X_s \, \mu^+ \mu^-$} 
and \bm{$B\to X_s \, \t^+ \t^-$} 
\end{center}
\vspace{0.05cm}
\begin{center}
\sc F.~Kr\"uger\footnote{\footnotesize
Electronic address: krueger@physik.rwth-aachen.de} and
L.\,M.~Sehgal\footnote{\footnotesize Electronic address: sehgal@physik.rwth-aachen.de}
\\ \it Institut f\"ur Theoretische Physik (E), RWTH Aachen\\
D-52074 Aachen, Germany
\end{center}
\vspace{1.0cm}
\thispagestyle{empty}
\centerline{\bf ABSTRACT}
\begin{quotation}
The effective \ha\ for the decay $b\to s\, l^+ l^-$ predicts a characteristic polarization for the final state
lepton, which can serve as an important test of the underlying theory. The lepton polarization has, in addition
to a longitudinal component $\lop$, two orthogonal components $\trp$ and $\nop$, lying in and perpendicular
to the decay plane which are proportional to $m_l/m_b$, and therefore significant for the $\t^+\t^-$ channel. 
The normal polarization component $\nop$ is a \T-odd effect connected with the nonhermiticity of the 
effective \ha, arising mainly from $c\bar{c}$ intermediate states. We calculate all three polarization components for 
the decay $B\to X_s \, \t^+ \t^-$ as a function of the lepton pair mass, and find average values 
$\left\langle \lop \right\rangle_{\t} = -0.37$, $\left\langle \trp \right\rangle_{\t} = -0.63$, $\left\langle \nop \right\rangle_{\t} = 0.03$.
By comparison, the $\mu^-$ polarization is $\left\langle \lop\right\rangle_{\m} = -0.77$, 
$\left\langle \trp \right\rangle_{\m} = \left\langle \nop \right\rangle_{\m} \approx 0$.
\end{quotation}
\begin{quote}
\end{quote}
\setcounter{footnote}{0}
%
%
\newpage
\Sec{INTRODUCTION} \label{intro}
\bigskip
The decay $B\to X_s \, l^+ l^-$ has received considerable attention as a potential testing ground for the 
effective \ha\ describing flavour-changing neutral current processes in $B$ decay (see e.g.~\rf{bdecay}). This
\ha\ contains the one-loop effects of the electroweak interaction, which are sensitive to the top-quark
mass \cite{topquark, dt, bm}. In addition, there are important QCD corrections 
\cite{gsw, cella, godsn, misiak}, which have recently been
calculated in next-to-leading order \cite{bm, misiakerra}. The inclusive distributions have been studied in 
\cite{gsw, amm, agm},
while the exclusive channels $B\to K l^+ l^-$ and $B\to K^* l^+ l^-$ have been analysed in 
\cite{dt, exclusive}.
Recently, attention has been drawn to the fact that the longitudinal polarization of the lepton, $\lop$, in
$B\to X_s \, l^+ l^-$ is an important observable, that may be especially accessible in the mode 
$B\to X_s \, \t^+ \t^-$ \cite{joanne}. The purpose of this paper is to show that complementary information is 
contained in the two orthogonal components of polarization ($\trp$, the component in the decay plane, and 
$\nop$, the component normal to the decay plane), both of which are proportional to $m_l/m_b$, and 
therefore significant for the $\t^+\t^-$ channel. The normal component $\nop$, in particular, is a novel 
feature, since it is a \T-odd observable, that comes about because of the nonhermiticity of the effective \ha, 
associated with real $c\bar{c}$ intermediate states.
\vspace{1.5cm}

\Sec{SHORT-DISTANCE CONTRIBUTIONS}
\bigskip
The effective short-distance \ha\ for $b\to s\, l^+l^-$ \cite{bm,gsw,godsn,misiak} leads to the QCD-corrected
matrix element
\bea\label{heff}
\cal{M}&=&\left.\frac{4 G_F}{\sqrt{2}}V_{tb}^{}V_{ts}^*\,\frac{\a}{4\p}\right\{\ceff \left( \bar{s}\g_{\m}
P_L b \right)
\bar{l}\g^{\m}l+ c_{10}\left( \bar{s}\g_{\m}P_L b \right)\bar{l}\g^{\m}\g^5 l\nnu\\
& &\mbox{}-2\,\cseff \bar{s}\left. i  \sigma_{\m\n}\frac{q^{\n}}{q^2}\left(m_b P_R + m_s P_L\right)b\, 
\bar{l}\g^{\m} l\right\} ,
\eea
where $P_{L,R} = \frac{1}{2} (1\mp\gamma_5)$, and the analytic expressions for the Wilson coefficients
are given in \rf{bm}. We use in our analysis the parameters given in the
Appendix and obtain in leading logarithmic approximation 
\be\label{wilsonc7c10}
\cseff = -0.315,\quad c_{10} = -4.642\ , 
\ee
and in next-to-leading order
\bea\label{wilsonc9}
\ceff&=&c_9  \tilde{\eta}(\sh) + g(\mch,\sh) \left(3 c_1 + c_ 2 +3c_3 + c_4 + 3c_5 + c_6\right) 
-\frac{1}{2} g(\msh,\sh) \left(c_3 + 3 c_4\right)\nnu\\
&& - \,\frac{1}{2} g(\mbh,\sh)\left(4c_3 + 4c_4 +3c_5+c_6\right)+ \frac{2}{9}\left(3c_3 +c_4+3c_5+c_6\right)\ ,
\eea
with
\bea\label{wilsonci}
\left(3 c_1 + c_ 2 +3c_3 + c_4 + 3c_5 + c_6\right) &=& 0.359\ ,\nnu \\ 
\left(4c_3 + 4c_4 +3c_5+c_6\right)& =& - 6.749\times 10^{-2}\ , \nnu\\
\left(3c_3 +c_4+3c_5+c_6\right)&=&-1.558 \times 10^{-3}\ ,\\
\left(c_3 + 3 c_4\right)&=&- 6.594\times 10^{-2}\ , \nnu\\
c_9  \tilde{\eta}(\sh)&=&4.227 + 0.124\ \omega(\sh)\ , \nnu
\eea
where we have introduced the notation $\sh = q^2/m_b^2$, $\mh_i = m_i/m_b $. The function $\omega(\sh)$
represents the one-gluon correction\bem{See Refs.~\cite{bm} and \cite{misiakerra}. Here we neglect
corrections due to a nonzero lepton mass. This will be discussed in a further publication.} to the 
matrix element of 
the operator ${\cal O}_9$, while $g(\mh_i, \sh)$ 
arises from the one-loop contributions of the four-quark operators ${\cal O}_1$ and ${\cal O}_2$, and is
given by 
\bea\label{loopfunc}
g(\mh_i,\sh)&=&-\frac{8}{9}\ln(\mh_i)+\frac{8}{27}+\frac{4}{9}y_i
-\frac{2}{9}(2+y_i)\sqrt{|1-y_i|}\nnu\\
&&\times\left\{
\Theta(1-y_i)(\ln\left(\frac{1 + \sqrt{1-y_i}}{1 - \sqrt{1-y_i}}\right)-i\p)
+ \Theta(y_i-1) 2\arctan\frac{1}{\sqrt{y_i-1}}\right\}\ ,\nnu\\
\eea
where $y_i \equiv 4 \mh_i^2/\sh$. 
\vspace{1.5cm}

\pagebreak
\Sec{LONG-DISTANCE CONTRIBUTIONS}
\bigskip
In addition to the short-distance interaction defined by \eqs{heff}{wilsonci} 
it is possible to take into account long-distance effects, associated with real $c\bar{c}$ resonances in the 
intermediate states, i.e.~with the reaction chain $B\to X_s + V(c\bar{c})\to X_s\, l^+ l^-$. This can be 
accomplished in an approximate manner through the substitution \cite{longdist}
\be\label{long-dist}
g(\mch,\sh) \longrightarrow g(\mch,\sh)-\frac{3\p}{\a^2}\sum\limits_{V= J/\psi, \psi', \dots}
\frac{\mh_V {\branch}(V\to l^+l^-)\hat{\G}_{\mathrm {total}}^V}
{\sh-\mh^2_V + i \mh_V \hat{\G}_{\mathrm {total}}^V}\ ,
\ee
where the properties of the vector mesons are summarized in Table \ref{table1}.
%
%
\begin{table}[h]
\begin{center}\caption{Charmonium ($c\bar{c}$) masses and widths \cite{pdg}.}\label{table1}
\vspace{0.3cm}
\begin{tabular} {c  c  c  c}
\hline\hline
Meson & Mass $(\GeV)$ & ${\branch}(V\to l^+l^-)$ & $\G_{\mathrm {total}}$ $(\MeV)$\\  \hline 
$ J/\Psi  $(1S) & 3.097& $6.0\times 10^{-2}$ & $0.088$\\
$ \Psi $(2S) & 3.686 & $8.3\times 10^{-3}$ & $0.277$\\
$ \Psi (3770) $ &3.770 &$1.1\times 10^{-5}$ &23.6 \\
$ \Psi (4040)$ & 4.040&$1.4\times 10^{-5}$ &52 \\
$ \Psi (4160)$ &4.159 & $1.0\times 10^{-5}$& 78\\
$ \Psi (4415) $ &4.415 &$1.1\times 10^{-5}$ &43 \\ \hline\hline
\end{tabular}
\end{center}
\end{table}
%
%
There are six known resonances in the $c\bar{c}$ system that can 
contribute to the decay modes $B\to X_s\,  e^+ e^-$ and $B\to X_s\, \mu^+\mu^-$. Of these, all except
the lowest $J/\psi (3097)$ contribute to the channel $B\to X_s\,  \t^+ \t^-$, for which the invariant mass
of the lepton pair is $s > 4m_{\t}^2$.\bem{As noted by several authors \cite{amm,lw}, the ansatz 
(\ref{long-dist}) for the resonance contribution implies an inclusive direct $J/\psi$ production rate 
${\branch}(B\to J/\psi X_s) = 0.15\%$ that is $\sim 5$ times smaller than the measured $J/\psi$ rate \cite{jpsi}.
This is usually amended by the introduction of a phenomenological factor $\kappa_V \approx 2$ multiplying
the Breit-Wigner function in (\ref{long-dist}). In the present paper, the only observable that is
significantly affected by this change is the polarization component $\nop$, where we will show the results
for both $\kappa_V = 1$ and $\kappa_V = 2.35$.}

Alternatively, we can express $g(\mh_i,\sh)$, \eq{loopfunc}, through the renormalized vacuum
polarization $\Pi^{\g}_{\mathrm{had}}(\sh)$, which is related to the experimentally measurable quantity 
$\rhadron(\sh)\equiv \sigma_{\mathrm{tot}}(e^+e^-\to{\mathrm{hadrons}})/\sigma(e^+e^-\to\m^+\m^-)$
via a dispersion relation \cite{dispersion}, i.e.
\be\label{vacpola}
\Re \Pi^{\g}_{\mathrm{had}}(\sh) = \frac{\a \sh}{3\p}\ P \int\limits_{4\hat{m}_{\p}^2}^{\infty}
\frac{R_{\mathrm {had}}(\sh')}{\sh'(\sh'-\sh)}d\sh',\quad {\mathrm{and}}\quad
\Im \Pi^{\g}_{\mathrm{had}}(\sh) = \frac{\a}{3} \rhadron(\sh)\ ,
\ee
where $P$ denotes the principal value.
Using these relations, one finds for the $c\bar{c}$ contribution
\be\label{Img}
\Im g(\mch,\sh) = \frac{\p}{3} \rhadroncc\ ,
\ee
and
\be\label{Reg}
\Re  g(\mch,\sh) = -\frac{8}{9} \ln \mch -\frac{4}{9} + \frac{\sh}{3}\ P \int\limits_{4\hat{m}_{D}^2}^{\infty}
\frac{R_{\mathrm {had}}^{c\bar{c}}(\sh')}{\sh'(\sh'-\sh)}d\sh'\ . 
\ee
The cross-section ratio $R_{\mathrm {had}}^{c\bar{c}}$ may be written as 
\be\label{Rhad}
\rhadroncc = \rcontcc + \rrescc \ ,
\ee
where $R_{\mathrm {cont}}^{c\bar{c}}$ and $R_{\mathrm {res}}^{c\bar{c}}$ denote the contributions 
from the continuum and the narrow resonances, respectively. 
The latter is given by the Breit-Wigner formula 
\be\label{BW}
\rrescc  =\sum\limits_{V= J/\psi, \psi', \dots}
 \frac{9\sh}{\a^2}\ \frac{{\branch}(V\to l^+l^-)\hat{\G}_{\mathrm {total}}^{V}
\hat{\G}_{\mathrm {had}}^{V}}
{(\sh-\mh^2_V)^2 +  \mh_V^2 \hat{\G}_{\mathrm {total}}^{V^2}}\ , 
\ee
whereas $R_{\mathrm {cont}}^{c\bar{c}}$ can be determined using the experimental data.
We use the parametrization of $R_{\mathrm {cont}}^{c\bar{c}}$ given in \cite{bp} (see Appendix).
In Fig.~\ref{figimg}, the imaginary part of $g(\mch,\sh)$, \eq{loopfunc}, is plotted and compared
with $\Im g(\mch, \sh)$, \eq{Img}. Our numerical results are based on Eqs.~(\ref{Img}) and (\ref{Reg}), 
with the parameter $\kappa_V$ chosen equal to $2.35$. 
\vspace{1.5cm}

\pagebreak
\Sec{RATE AND FORWARD-BACKWARD ASYMMETRY}
\bigskip
Neglecting nonperturbative corrections \cite{fls}, the decay width as a function of the invariant mass of 
the lepton pair ($q^2\equiv m^2_{l^+ l^-}$) is given by 
\be\label{rate}
\frac{d\G}{d\sh} = \frac{G_F^2 m_b^5}{192 \p^3}\,\frac{\a^2}{4\p^2}\,|V_{tb}^{}V_{ts}^*|^2\,
\l^{1/2}(1,\sh,\msh^2) \,\sqrt{1-\frac{4 \mlh^2}{\sh}}\,\Delta\ ,
\ee
where the factors $\lambda$ and $\Delta$ are defined by
\be\label{kinfunc}
\l (a,b,c) = a^2 + b^2 + c^2 - 2 (a b + b c + a c)\ ,
\ee
and
\bea\label{delta}
\Delta &=& \Bigg\{\left(12\, \Re(\cseff \ceff)F_1(\sh,\msh^2) + \frac{4}{\sh}|\cseff|^2 F_2(\sh,\msh^2)\right)
\left(1+\frac{2\mlh^2}{\sh}\right)\nnu\\
&&+\left(|\ceff|^2 + |c_{10}|^2\right) F_3(\sh,\msh^2,\mlh^2)+ 6\, \mlh^2\left(|\ceff|^2 
-|c_{10}|^2\right) F_4(\sh,\msh^2)\Bigg\}\ ,\nnu\\
\eea
with
\bea\label{kinfuncs}
F_1(\sh,\msh^2)& =&(1-\msh^2)^2 - \sh(1+\msh^2)\ ,\\
F_2(\sh,\msh^2)&=& 2 (1+\msh^2) (1-\msh^2)^2 -\sh (1 + 14 \msh^2 +\msh^4)-\sh^2 (1+\msh^2)\ ,\nnu\\
\\
F_3(\sh,\msh^2,\mlh^2) &=& (1-\msh^2)^2 +\sh (1+\msh^2) -2 \sh^2 
+ \l(1,\sh,\msh^2) \frac{2\mlh^2}{\sh}\ ,\\
F_4(\sh,\msh^2)&=&1- \sh +\msh^2\ .
\eea
If the lepton mass $m_l$ is neglected, we recover the results of Ali \ea \cite{amm,agm}. 
If the strange-quark mass 
$m_s$ is also neglected, we obtain the results of Grinstein \ea \cite{gsw} and Buras and M\"unz \cite{bm}.
Finally, for $m_l \neq 0$ but $m_s = 0$, we reproduce the result given by Hewett \cite{joanne}.

To complete the correspondence with previous results, we give here the forward-backward asymmetry of
the lepton $l^-$ in the $l^+ l^-$ centre-of-mass system:
\be\label{asymmetry}
\fba = -3\, \l^{1/2}(1,\sh,\msh^2) \,\sqrt{1-\frac{4 \mlh^2}{\sh}} \ \frac{c_{10}\left[ \sh\,\Re \ceff 
+ 2\,\cseff(1+\msh^2)\right]}{\Delta}\ .
\ee
This agrees with \rf{amm} when $m_l$ is neglected.
\vspace{1.5cm}

\Sec{LEPTON-POLARIZATION}
\bigskip
We now proceed to a discussion of the inclusive lepton polarization. We define three orthogonal unit vectors
\bea\label{unitvec}
\eL&=& \frac{\Vec{p}_-}{|\Vec{p}_-|}\ , \nnu \\
\eN&=&(\Vec{p}_s\times\Vec{p}_-)/|\Vec{p}_s\times\Vec{p}_-|\ ,\\
\eT&=&  \eN\times \eL\nnu\ ,
\eea 
where $\Vec{p}_-$ and $\Vec{p}_s$ are three-momenta of the $l^-$ and the $s$ quark, respectively, in the
c.m.~frame of the $l^+ l^-$ system. The differential decay rate of $B\to X_s \, l^+ l^-$ for any given spin
direction $\Vec{n}$ of the lepton $l^-$, where $\Vec{n}$ is a unit vector in the $l^-$ rest frame, may then 
be written as
\be\label{spinrate}
\frac{d\G(\Vec{n})}{d\sh} = \frac{1}{2}\left.\left.\left(\frac{d\G}{d\sh}\right)_{\mathrm{unpol} }\right[1 + 
\left(\lop \eL + \trp \eT + \nop\eN\right)\cdot\Vec{n}\right] ,
\ee
where $\lop$, $\trp$, $\nop$ are functions of $\sh$, which give the longitudinal, transverse and normal
components of polarization. The polarization component $P_i$ $( i=\mathrm {L, T, N})$ is obtained by
evaluating
\be\label{pola}
P_i(\sh) = \frac{d\G(\Vec{n}=\Vec{e}_i)/d\sh -d\G(\Vec{n}=- \Vec{e}_i)/d\sh}{ d\G(\Vec{n}=\Vec{e}_i)/d\sh 
+d\G(\Vec{n}=-\Vec{e}_i)/d\sh}\ .
\ee

Our results for the polarization components $\lop$, $\trp$ and $\nop$ are as follows
\bea
\lop(\sh) &=& \sqrt{1-\frac{4\mlh^2}{\sh}}\left [12\,\cseff c_{10}\left((1-\msh^2)^2-\sh(1+\msh^2)\right)
\right.\nnu\label{polaL}\\
&&+\left.2\,\Re(\ceff c_{10})\left((1-\msh^2)^2 + \sh (1+\msh^2) - 2\sh^2\right)\right]/\Delta\ , \\\nnu\\
\trp(\sh)&=&\frac{3\p \mlh}{2\sqrt{\sh}}\l^{1/2}(1,\sh,\msh^2) \left[\cseff c_{10} (1-\msh^2) 
-4\,\Re (\cseff\ceff)(1+\msh^2)\right.\nnu\\ 
&&-\frac{4}{\sh}\left.|\cseff|^2 (1-\msh^2)^2 + \Re(\ceff c_{10}) (1-\msh^2)-
|\ceff|^2\sh\right]/\Delta\ , \\\nnu\\
\nop(\sh) &=& \frac{3\p \mlh}{2 \Delta} \Im(\c9eff* c_{10}) \sqrt{\sh}\,\l^{1/2}(1,\sh,\msh^2)\,
\sqrt{1-\frac{4\mlh^2}{\sh}}\ .
\eea 

The expression for $\lop$ agrees with that obtained by Hewett \cite{joanne}, when we set $\msh = 0$ in 
\eq{polaL}.

It should be noted that the polarization components $\lop$, $\trp$ and $\nop$ involve different quadratic
functions of the effective couplings $\cseff$, $\ceff$ and $c_{10}$, and therefore contain independent
information. The component $\nop$ is proportional to the absorptive part of the effective coupling $\ceff$,
which is dominated by the charm-quark contribution (cf.~\eq{wilsonc9}).
It is obvious that the polarization is affected by any alteration in the relative magnitude and sign of 
$\cseff$, $\ceff$ and $c_{10}$, and thus can serve as a probe of possible new interactions transcending 
the standard model.

The polarization components $\lop$, $\trp$ and $\nop$ are plotted in \figs{figLmu}{figN}. In the case of the
$\m^+\m^-$ channel, the only significant component is $\lop$, which has a large negative value over most
of the $\sh$ domain, with an average value $\left\langle \lop \right\rangle_{\m} = -0.77$. By contrast, all three 
components are sizeable in the $\t^+\t^-$ channel, with average values $\left\langle \lop \right\rangle_{\t} 
= -0.37$, $\left\langle \trp \right\rangle_{\t} = -0.63$, $\left\langle \nop \right\rangle_{\t} = 0.03\,(0.02)$ for
$\kappa_V= 2.35\,(1)$. Notice that the
\T-odd component $\nop$, though small, is considerably larger than the corresponding normal
polarization of leptons in $K_L\to\pi^+\m^-\bar{\n}$ or $K^+\to\pi^+\m^+\m^-$, which requires a final
state Coulomb interaction of the lepton with the other charged particles, and is typically of order
$\a (m_{\m}/m_K)  \sim 10^{-3}$ \cite{kdecay}.

The inclusive branching ratios are predicted to be ${\branch}(B\to X_s \, \m^+ \m^-) = 6.7\times 10^{-6}$, 
${\branch}(B\to X_s \, \t^+ \t^-) = 2.5\times 10^{-7}$. 
The lower rate of the $\t^+\t^-$ channel may be offset by the fact
that the decay of the $\t$ acts as a self-analyser of the $\t$ polarization. 
Assuming (as in \rf{joanne}) a total of $5\times 10^8$ $B\bar{B}$ decays, one can expect to observe
$\sim 100$ identified $B\to X_s \, \t^+ \t^-$ events, permitting a test of the predicted polarizations
$\left\langle \lop \right\rangle = -0.37$, and $\left\langle \trp \right\rangle = -0.63$ with good accuracy. 
We shall discuss in a more detailed
paper the dependence of the lepton polarization on the production angle $\theta$, the spin correlation of
the $l^+ l^-$ pair, the influence of nonperturbative effects (quark-binding corrections), as well as
lepton-spin effects in exclusive channels.
\vspace{1.5cm}

\centerline{\bf ACKNOWLEDGEMENTS}
We would like to thank H. Burkhardt for providing us with the parametrization of  $\rcont$.
One of us (F. K.) is indebted to the Deutsche Forschungsgemeinschaft (DFG) for the award of a Doctoral 
stipend.
\vspace{1.5cm}

\app{INPUT PARAMETERS}

\label{input}
\parbox{13cm}{
\bea
&&m_b=4.8\ \GeV,\ m_c= 1.4\ \GeV,\ m_s=0.2\ \GeV,
\ m_t=176\ \GeV\ ,\nnu\\
&&m_{\m} = 0.106\ \GeV,\ m_{\t} = 1.777\ \GeV,\ M_W=80.2\ \GeV,\ \m=m_b\ , \nnu\\
&&V_{tb}^{}=1,\ V_{ts}^{*}= -V_{cb}^{},\ {\branch} (B\to X_c\,l\bar{\n}_l)= 10.4\%\ , \nnu \\
&&\Lambda_{\mathrm{QCD}} = 225\ \MeV,\ \a=1/129,\ {\sin}^2{\theta_{\mathrm{W}}} = 0.23\ .
\footnotemark\nnu
\eea}\hfill\parbox{1.3cm}{\bea\eea}
\footnotetext{We use $\a_s(\m)$ that is given by the formula (4.12) of \rf{bm}.}
\bigskip
\bea
\rcontcc = 
\left\{\begin{array}{l} 0 \quad{\mathrm{for}}\quad 0\leq \sh \leq 0.60\ ,\\ 
-6.80  + 11.33 \sh\quad{\mathrm{for}}\quad 0.60\leq \sh \leq 0.69\ ,\\
1.02\quad{\mathrm{for}}\quad 0.69 \leq \sh \leq 1\ .
\end{array}\right.
\eea
%
%

%
\newpage
\centerline{\bf FIGURE CAPTIONS}
\begin{enumerate}
\item[\bf Figure 1] The imaginary part $\Im g(\mch, \sh)$ as a function of the invariant mass of the lepton pair.
The dashed line represents the theoretical result, neglecting long-distance effects, and the solid curve
shows the imaginary part using $\rhadroncc$, as described in the text.
\item[\bf Figure 2] The longitudinal polarization $\lop$ for the $\m^-$ including the $c\bar{c}$ 
resonances ($\kappa_V = 2.35$).
\item[\bf Figure 3] The longitudinal polarization $\lop$ for the tau lepton including $c\bar{c}$
resonances.
\item[\bf Figure 4] The transverse polarization $\trp$ (in the decay plane) of the $\t^-$ for $\sh \geq 4 \mth^2$.
\item[\bf Figure 5] The normal polarization $\nop$, i.e.~normal to the decay plane, in the $\t^+\t^-$ channel
including $c\bar{c}$ intermediate states. The solid line corresponds to $\kappa_V = 2.35$, the
dashed one is $\kappa_V = 1$.
\end{enumerate}
%
%
\newpage
\begin{figure}[h]
\centerline{\epsfysize=20cm\epsffile{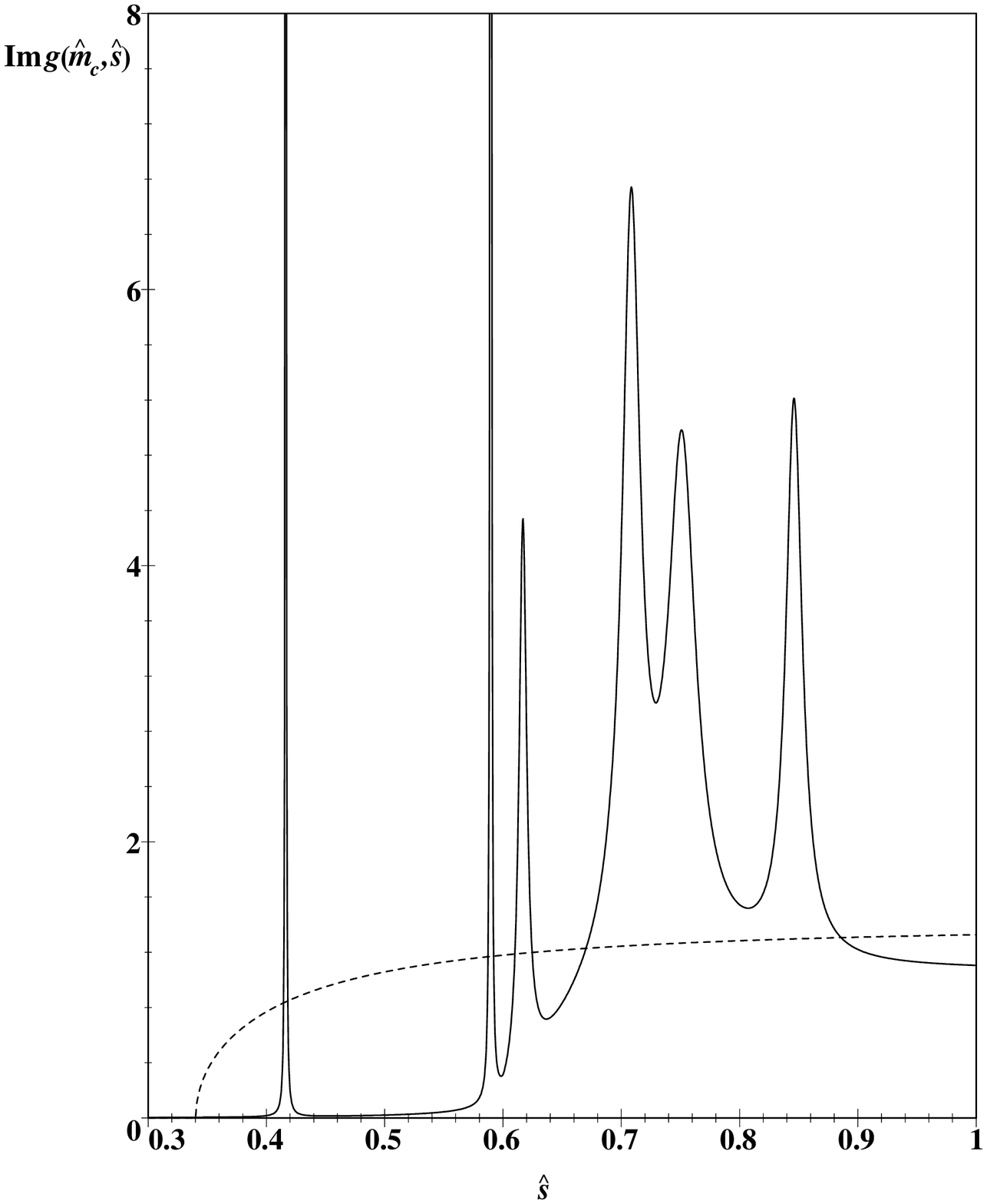}}
\caption{}\label{figimg}
\end{figure}
%
%
\begin{figure}[h]
\centerline{\epsfysize=20cm\epsffile{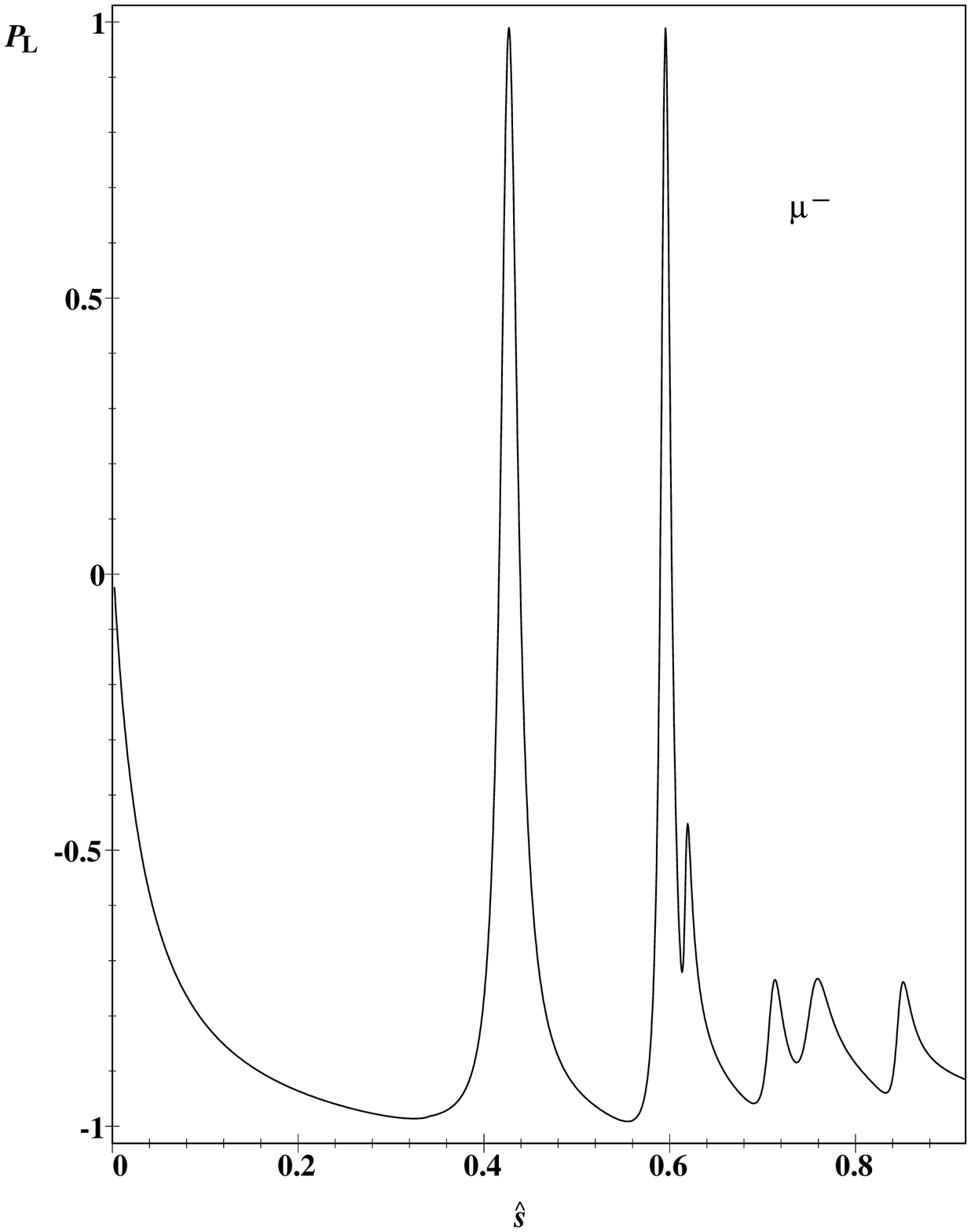}}
\caption{}\label{figLmu}
\end{figure}
%
%
\begin{figure}[h]
\centerline{\epsfysize=20cm\epsffile{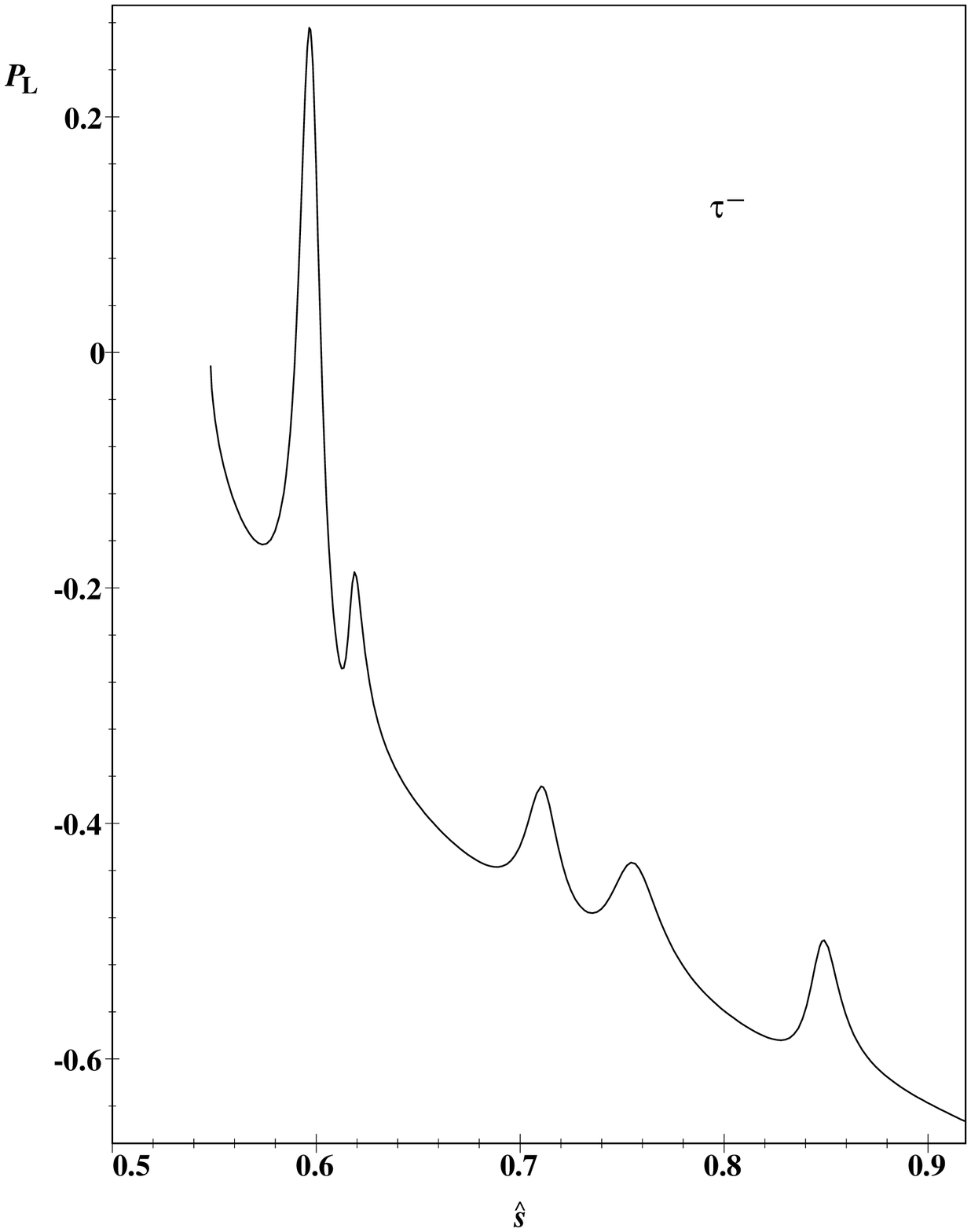}}
\caption{}\label{figL}
\end{figure}
%
%
\begin{figure}[h]
\centerline{\epsfysize=20cm\epsffile{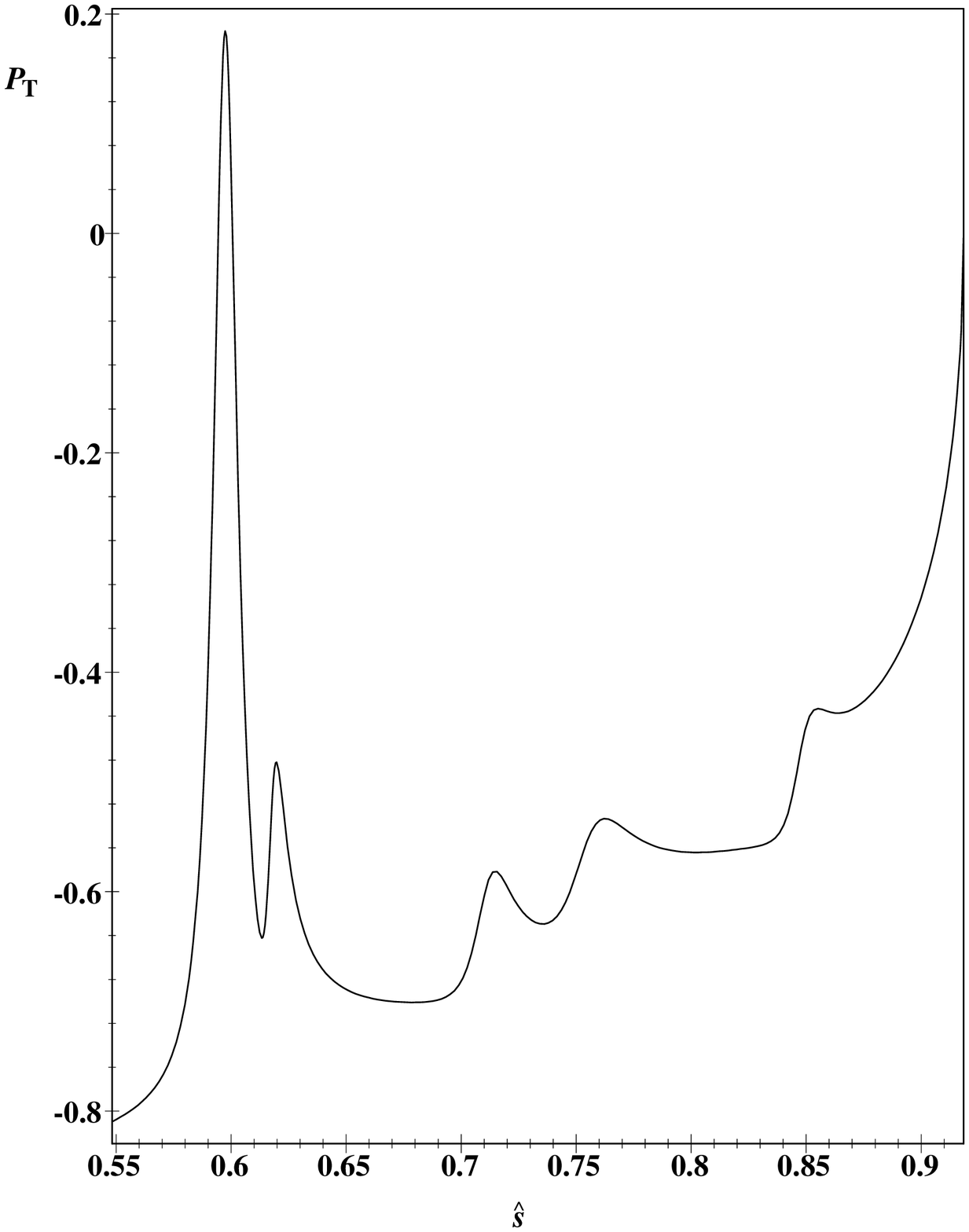}}
\caption{}\label{figT}
\end{figure}
%
%
\begin{figure}[h]
\centerline{\epsfysize=20cm\epsffile{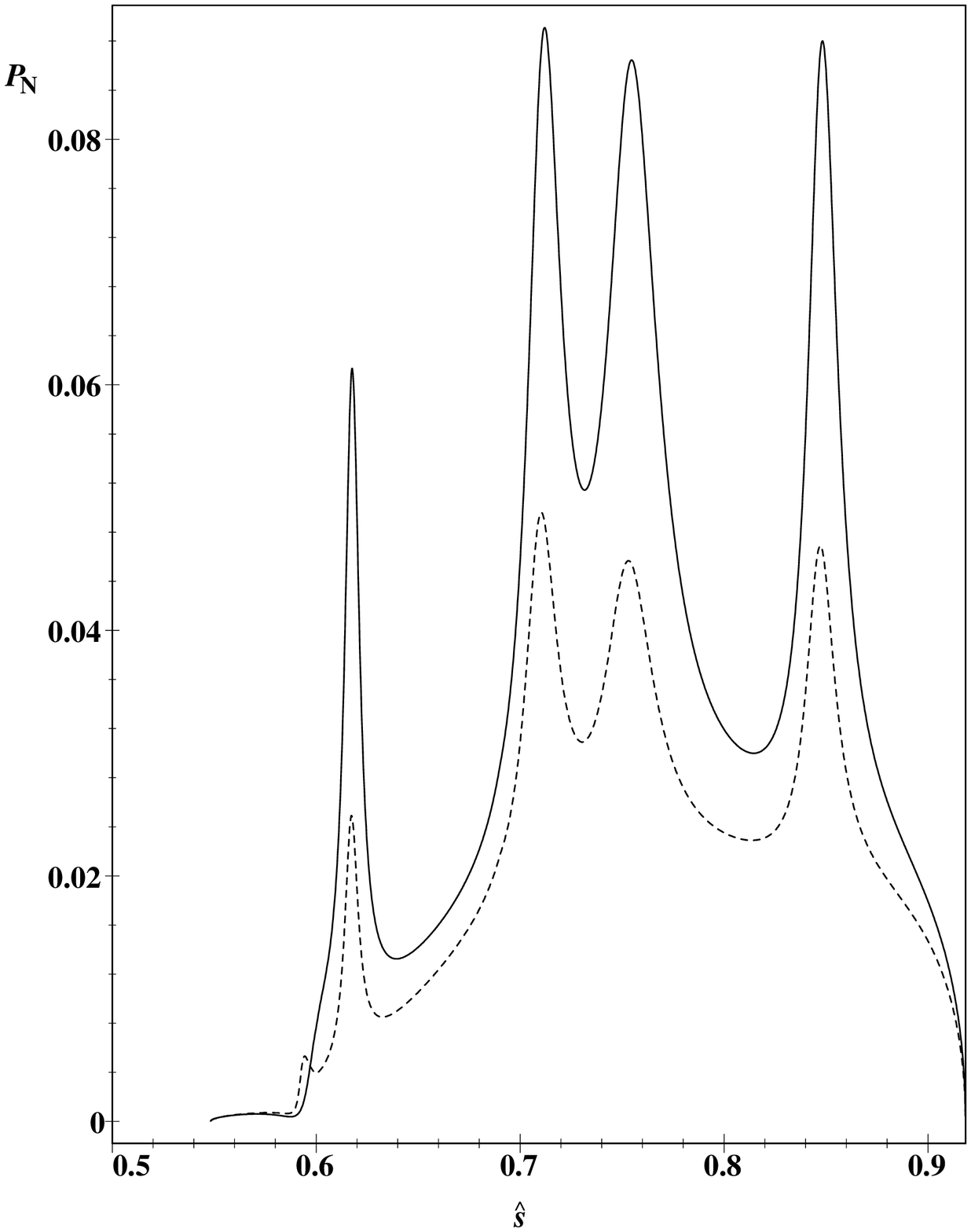}}
\caption{}\label{figN}
\end{figure}
%
%

\end{document}